\documentclass[letterpaper,titlepage,11pt]{article}

\pdfoutput=1

\usepackage{amssymb,amsmath,amsfonts}
\usepackage{epsfig}
\usepackage{ccaption}
\usepackage{graphicx}

\usepackage[
      colorlinks=true,
      linkcolor=blue,
      urlcolor=blue,
      filecolor=black,
      citecolor=red,
      pdfstartview=FitV,
      pdftitle={},
        pdfauthor={Roberto Emparan, Alejandro Fern\'andez-Piqu\'e, Raimon Luna},
        pdfsubject={},
        pdfkeywords={},
        pdfpagemode=None,
        bookmarksopen=true
      ]{hyperref}

\def\clock{{\count0=\time
           \divide\count0 60
           \ifnum\count0<10 0\fi\the\count0
           \multiply\count0 -60 \advance\count0 \time
           :\ifnum\count0<10 0\fi \the\count0
         }}
\newcommand{\timestamp}{{\small\vbox{\hbox{\tt\jobname.tex}
\hbox{\the\day/\the\month/\the\year, \clock}}}}

\newcommand{\ie}{{\it i.e.,\,}}
\newcommand{\eg}{{\it e.g.,\,}}

\newcommand{\lp}{\left(}
\newcommand{\rp}{\right)}

\newcommand{\beq}{\begin{equation}}
\newcommand{\eeq}{\end{equation}}
\newcommand{\bea}{\begin{eqnarray}}
\newcommand{\eea}{\end{eqnarray}}
\newcommand{\beqa}{\begin{eqnarray}}
\newcommand{\eeqa}{\end{eqnarray}}

\numberwithin{equation}{section}

\begin{document}

\begin{titlepage}
\leftline{}
\vskip 2cm
\centerline{\LARGE \bf Geometric polarization of plasmas and} 
\bigskip
\centerline{\LARGE \bf Love numbers of AdS black branes}
\vskip 1.2cm
\centerline{\bf Roberto Emparan$^{a,b}$, Alejandro Fern\'andez-Piqu\'e$^{b}$, Raimon Luna$^{b}$}
\vskip 0.5cm
\centerline{\sl $^{a}$Instituci\'o Catalana de Recerca i Estudis
Avan\c cats (ICREA)}
\centerline{\sl Passeig Llu\'{\i}s Companys 23, E-08010 Barcelona, Spain}
\smallskip
\centerline{\sl $^{b}$Departament de F{\'\i}sica Qu\`antica i Astrof\'{\i}sica, Institut de
Ci\`encies del Cosmos,}
\centerline{\sl  Universitat de
Barcelona, Mart\'{\i} i Franqu\`es 1, E-08028 Barcelona, Spain}
\smallskip

\vskip 1.2cm
\centerline{\bf Abstract} \vskip 0.2cm 
\noindent 
We use AdS/CFT holography to study how a strongly-coupled plasma polarizes when the geometry where it resides is not flat. We compute the linear-response polarization coefficients, which are directly related to the static two-point correlation function of the stress-energy tensor. In the gravitational dual description, these parameters correspond to the tidal deformation coefficients---the Love numbers---of a black brane. We also compute the coefficients of static electric polarization of the plasma.

\end{titlepage}
\pagestyle{empty}
\small
\normalsize
\newpage
\pagestyle{plain}
\setcounter{page}{1}

\section{Introduction}

When a continuous system is placed in a generic curved geometry (which we will always assume is time-independent), it polarizes: its energy density, pressure, and other components of the stress-energy tensor acquire inhomogeneous expectation values. This happens in any state of the system, in particular in the vacuum ground state and in finite-temperature plasmas. 

When the deformation of the geometry away from flatness is small, this polarizability is captured by a set of linear-response coefficients. These are determined by the two-point correlation function of the stress-energy tensor, and hence they carry non-trivial information about the system. They are experimentally accessible data, especially for systems in two space dimensions, where the background geometry can be more easily manipulated. It is therefore of interest to have theoretical computations of their values. 

For systems that are strongly coupled the best available tool for these calculations is the AdS/CFT correspondence, where one solves a dual, weakly coupled gravitational system. In the case of interest to us here---the geometric polarization of a conformally invariant, finite-temperature plasma---we will study how the gravitational dual to the thermal state, namely an Anti-deSitter black brane, gets distorted when the boundary geometry is changed from flat Minkowski spacetime to a generic, weakly-deformed time-independent geometry. This amounts to introducing a static gravitational potential at infinity, which we may think of as an external gravitational source that induces a tidal deformation of the black brane. In order to compute this deformation, we solve the equations for a linearized perturbation of the geometry that satisfies an appropriate boundary condition at infinity. Namely, the metric perturbation must not vanish asymptotically, but instead approach the non-zero value that matches the source, \ie\ the metric perturbation specified at the boundary.

Interestingly, the static tidal polarization is also of relevance in the field of black hole and stellar astrophysics. There, the linear-response coefficients are known as the Love numbers of the gravitating object \cite{Damour:2009vw,Binnington:2009bb}. The asymptotic external sources are a way of approximating the effect of other distant massive bodies which pull gravitationally on the object, and the Love numbers characterize its response. In recent times these Love numbers have been a matter of interest (and of corny puns) in this area, since they may be measured from the gravitational wave signal of inspiralling black holes and neutron stars \cite{Flanagan:2007ix,Hinderer:2007mb}. One can then use them to test the predictions of General Relativity, and also to extract information about the internal constitution and equation of state of neutron stars.

Given that Love numbers are eminently measurable quantities, not only in astrophysics but also in the AdS/CMT correspondence, it may be surprising that---to the best of our knowledge---they have not been explicitly investigated in the latter context. We shall do so in this article. Specifically, we will compute the Love numbers for black branes in AdS$_4$ and in AdS$_5$. The former case is dual to the response of a strongly-coupled plasma to the curvature of the 2+1-dimensional geometry it resides in. This is presumably closer to physical realization in the lab (with all the caveats that attend to AdS/CMT modelling) than in the case of the 3+1-dimensional quark-gluon plasma dual to AdS$_5$. We will also show how the two-point function of the stress-energy tensor is obtained from the Love numbers \cite{Lennon:1967}.

Admittedly, this is not the first study of AdS black branes tidally polarized by an external gravitational source. Previous constructions of black branes spatially modulated by curved boundary geometries include \cite{Donos:2012js,Donos:2014uba,Donos:2014oha,Donos:2014gya}. These sources have been introduced in AdS/CMT with the primary aim of mimicking the breaking of translation invariance by an ionic lattice, so that, subsequently, phenomena like momentum relaxation can be studied. That is, here the polarization of the brane is a convenient means towards a further effect.

Our motivation is different. For us, the inhomogeneity is not intended to model a discrete lattice structure, but rather it is a distortion of a background geometry that is essentially smooth (on long enough scales), and whose direct effects on the plasma are an object of study in themselves. In this respect, our analysis is perhaps closer in spirit to the holographic studies of CFTs in black hole backgrounds \cite{Marolf:2013ioa}, although our approach and aims are different.

Furthermore, at the technical level, in these previous works the deformations have been considered fully non-linearly, which is rather more difficult than our linear perturbation analysis. The former requires either the numerical solution of partial differential equations, or instead very special deformations with a large degree of symmetry that renders them more solvable, but possibly less realistic. The linear-response analysis that we perform here has the advantage that it can be done simply and very generally. Surprisingly often, a linear approximation turns out to work well even for relatively large deformations, so perhaps this will also be true of our results. 
At any rate, in none of the previous studies are we aware of any attempt to compute the linear-response polarization coefficients of the vacuum and the plasma states. These are of enough interest to motivate our study.


The Love numbers---the polarization coefficients---are extracted from the solutions of linear, ordinary differential equations for gravitational perturbations of AdS black branes (which are the static limit of the equations in \cite{Kovtun:2005ev}). 
For the vacuum state, dual to empty AdS, we can solve the perturbation equations exactly and find the Love numbers in closed analytic form. For the thermal state, we will obtain them numerically and also analytically in a hydrodynamic gradient expansion, \ie\ in power series of the wavenumber $k$. A non-linear calculation of the hydrodynamic response of the plasma to an external gravitational force was made in \cite{Bhattacharyya:2008ji,Bhattacharyya:2008mz}, to lowest order for small $k$. When this result is applied to linearized deformations, we find agreement with our calculations. But by considering small amplitudes of the deformation, we can obtain results that extend to higher values of $k$.

Finally, as a natural extension of our study, we also compute the linear-response coefficients of the electric polarizability of the plasma.\footnote{The literature on AdS black branes deformed by boundary electric fields is too large, and more importantly, too differently motivated than ours, to properly refer to all of it here. We shall mention, though, that a linearized perturbation analysis was performed in \cite{Chesler:2013qla}.}

\section{Set up}

The solution for a neutral black brane in AdS$_{n+1}$ with cosmological constant $\Lambda=-\frac{n(n-1)}{2R^2}$ is
\beq
\frac{ds^2}{R^2}= \frac{dv^2}{v^2 f} + \frac{1}{v^2}\left( \eta_{\alpha\beta}+(1-f)u_\alpha u_\beta \rp dx^\alpha dx^\beta
\,,
\eeq
where $\alpha,\beta=1,\dots, n$ label the field theory directions, $u_\alpha$ is a timelike vector with $u_\alpha u_\beta \eta^{\alpha\beta}=-1$, and
\beq
f=1-\mu v^n\,.
\eeq
We denote the bulk radial coordinate as $v$, such that $v=0$ corresponds to the AdS boundary and $v=\mu^{-1/n}$ to the black brane horizon. The parameter $\mu$ determines the temperature $T$ of the configuration through
\beq
\mu= \lp\frac{4\pi T}{n}\rp^n\,.
\eeq
When $\mu\neq 0$ one can set $\mu=1$ without loss of generality. However, for the most part we will keep $\mu$ explicitly in our equations so we can easily recover the AdS vacuum by setting $\mu=0$.

The renormalized boundary metric 
\beq\label{renmet}
\gamma_{\alpha\beta}=\lim_{v\to 0}\frac{v^2}{R^2} g_{\alpha\beta}\eeq
in which the dual field theory lives is the flat Minkowski metric $\eta_{\alpha\beta}$. We want to study the response of the field theory to a small deformation of this geometry, which we decompose into plane waves,
\beq\label{pertgamma}
\gamma_{\alpha\beta}=\eta_{\alpha\beta}+ \bar{h}_{\alpha\beta} e^{i k_\alpha x^\alpha}\,.
\eeq
The $\bar{h}_{\alpha\beta}$ are constant numbers that characterize the relative amplitudes of the different metric deformations. We study time-independent perturbations, \ie\ with zero frequency
\beq
u^\alpha k_\alpha=0\,.
\eeq
This means that the perturbations are stationary, but not necessarily static since we allow non-zero components $u^\alpha\bar{h}_{\alpha\beta}$, which include momentum. We also allow non-zero gravitational potentials $u^\alpha u^\beta\bar{h}_{\alpha\beta}$.

Henceforth we partially fix the frame by choosing a time direction $t$ and aligning $k_\alpha$ with a direction $z$, \ie\
\beq
u^\alpha =\delta^\alpha{}_{t}\,,\qquad k_\alpha=k \,\delta_{\alpha z}\,.
\eeq
where $k$ is the wavenumber of the perturbation. 

In the gravitational problem we study small deformations of the black brane geometry that satisfy the Einstein-AdS equations. Fixing a radial gauge where $g_{vv}$ and $g_{v\alpha}$ remain unchanged,\footnote{With this, after requiring regularity of the geometry, the horizon position remains at the pole of $g_{vv}$ at $v=\mu^{-1/n}$.} the metric is perturbed as
\beq 
\frac{ds^2}{R^2}= \frac{dv^2}{v^2 f} + \frac{1}{v^2}\left( -f dt^2 + dz^2+\delta_{ij}dx^i dx^j +h_{\alpha\beta}(v) e^{ik z}d  x^\alpha dx^\beta\right)\,,
\eeq
where  $i,j=1,\dots,n-2$ label the coordinates $x^i$ orthogonal to $z$.

Near the asymptotic boundary we require that \eqref{pertgamma} holds, so
\beq
\lim_{v\to 0}h_{\alpha\beta}(v)= \bar{h}_{\alpha\beta}\,.
\eeq
Then the $\bar{h}_{\alpha\beta}$ are interpreted as asymptotic gravitational potentials acting on the black brane. 

\subsection{Gauge invariant perturbation analysis}

Following \cite{Kovtun:2005ev} we decompose the perturbations into scalars, vectors and tensors with respect to the group $O(n-2)$ of rotations orthogonal to the $z$ axis (the boost symmetries are broken at finite temperature).
In each of these channels one can find master variables $Z_{S,V,T}(v)$, in terms of which all the other metric components can be recovered, up to gauge transformations of the form $h_{\alpha\beta}\to h_{\alpha\beta}- 2\nabla_{(\alpha}\xi_{\beta)}$, with $\xi_\alpha=\xi_\alpha(v)e^{ikz}$, which leave the $Z$ invariant. Since the equations are linear and we want the perturbation to be non-zero at the boundary we can fix the normalization to
\beq\label{Zbdry}
Z_{S,V,T}(0)=1\,.
\eeq

For tensors and vectors the metric perturbations and the master variables are simply related,
\beqa
h_{ij}(v)&=&\bar{h}_{ij}^T\, Z_T(v)\label{ZTdef}\,,\\
h_{ti}(v)&=&\bar{h}_{ti}\, Z_V(v)\label{ZVdef}\,\,,
\eeqa
with $\bar{h}_{ij}^T$ a constant symmetric traceless tensor and $\bar{h}_{ti}$ a constant vector. 
For scalars the relation is 
\beq\label{ZSdef}
h_{tt}(v)+\frac12 \lp\frac{n}{n-2}-f\rp h(v)=\bar{H} Z_S(v)\,,
\eeq
where 
\beq
h(v)=\delta^{ij}h_{ij}(v)\,.
\eeq
At the boundary, \eqref{ZSdef} gives
\beq\label{Hbar}
\bar{H}=\bar{h}_{tt}+\frac1{n-2}\bar{h}\,.
\eeq

The tensor perturbations correspond to shearing deformations of the background geometry in planes orthogonal to $z$, which then induce shear in the plasma. The vectors create a stationary motion in the background, which will drag with it the black brane and impart momentum to the dual plasma\footnote{This motion creates vorticity in the plane $(x^i,z)$.}. The scalars introduce gravitational wells $\bar{h}_{tt}$ and averaged external  pressures $\delta^{ij}\bar{h}_{ij}$, which cause inhomogeneities in the energy density and local pressure of the plasma. 

From the Einstein equations in the bulk we derive the equations for the master tensor variable,\footnote{For $n=4$ these are the zero-frequency limit of the equations presented in \cite{Kovtun:2005ev}.}
\begin{equation}
Z_T''(v)-\frac{n-f }{f v}Z_T'(v)-\frac{k^2}{f} Z_T(v)=0\,,
\end{equation}
vector,
\begin{equation}
Z_V''(v)-\frac{n-1}{v}Z_V'(v)-\frac{k^2}{f}Z_V(v)=0\,,
\end{equation}
and scalar,
\begin{equation}\begin{split}
Z_S''(v)&+\frac{1}{v}\left(1-n\frac{(2f-1) (n-2)f+n}{ ( (n-2)f+n)f}\right) Z_S'(v)\\&+\frac{1}{f}\left(\frac{(1-f)^2 (n-2) n^2}{((n-2)f+n) v^2}-k^2\right) Z_S(v)=0.
\end{split}\end{equation}

Once $Z_S$ is obtained, the metric components in the scalar sector can be recovered using \eqref{ZSdef} and solving the first-order constraint  equations
\begin{equation}\label{hconst}
h'(v)=\frac{n(1-f)}{2 f^2 v}h_{tt}(v)+\frac{1}{f}h_{tt}'(v),
\end{equation}
and
\begin{equation}\label{hzzconst}
\begin{split}
h_{zz}'(v)=&\frac{ n (1-f)( (3 n-2)f-n)+4 f k^2 v^2}{2 f^2 v ( (n-2)f+n)}h_{tt}(v)\\&+
\frac{n(f-1)}{ ( (n-2)f+n)f}h_{tt}'(v)-\frac{2 k^2 v}{ (n-2)f+n}h(v).
\end{split}
\end{equation}

All the components of the metric perturbation that do not appear here can be gauge-fixed to zero. The component $h_{zz}(v)$ is partly constrained by the choice of radial gauge, but since the constraint \eqref{hzzconst} contains $h_{zz}'$ but not $h_{zz}$ there remains gauge freedom to always set 
\beq
\bar{h}_{zz}=0\,.
\eeq
In the boundary geometry this is simply achieved by changing $z\to z + c_z\, e^{ikz}$ with a suitable constant $c_z=O(\bar{h}_{\alpha\beta})$. 

Of all the other boundary values in the scalar sector, only $\bar{H}$ \eqref{Hbar} is physically meaningful, while $\bar{h}_{tt}$ and $\bar{h}$ separately are not. A Weyl transformation of the boundary geometry leaves $\bar{H}$ invariant, but changes $\bar{h}_{tt}$ and $\bar{h}$ separately. Thus the dual conformal field theory is only sensitive to $\bar{H}$.

This can also be understood from the bulk viewpoint. The functions $h_{tt}(v)$ and $h(v)$ are modified by bulk coordinate changes. In particular, a residual radial gauge transformation of the form
\beq\label{vchange}
v\to v\lp 1+ \frac{c_v}2 e^{ikz}\sqrt{1-\mu v^n}\rp
\eeq
with constant $c_v$ preserves the radial gauge condition at all $v$, and transforms
\beq\begin{split}\label{hvchange}
&h_{tt}(v)\to h_{tt}(v)+\frac{c_v}2 \lp n-(n-2)f\rp\sqrt{1-\mu v^n}\,,\\
&h(v)\to h(v)-c_v (n-2)  \sqrt{1-\mu v^n}\,,
\end{split}\eeq	
while $Z_S(v)$ and $\bar{H}$ remain invariant.\footnote{$h_{zz}(v)$ also changes, and keeping $h_{zv}=0$ requires an additional transformation $z\to z + \xi_z(v)e^{ikz}$.} One can now choose $c_v$ so that only $\bar{H}$, and not $\bar{h}$ nor $\bar{h}_{tt}$ separately, appears in the perturbed metric. This reflects the fact that changes in bulk radial gauge result into Weyl transformations at the boundary.

In this manner we can get rid of $\bar{h}_{tt}$ or $\bar{h}$ (insofar as they do not enter through $\bar{H}$), but one should be aware that the transformation \eqref{vchange} is not analytic near the horizon and generates terms in the metric of the form $\sim \sqrt{1-\mu v^n}$. A gauge where the metric components $h_{\alpha\beta}(v)$ are analytic on the horizon may be preferable over other gauges. In our subsequent calculations we will compute the values of $\bar{h}$ and $\bar{h}_{tt}$ that correspond to this analytic gauge. How this is done will be well illustrated with the hydrodynamic solution to the equations that we present in appendix~\ref{app:hydro}.
Bear in mind, however, that this is just a convenience: choosing the analytic gauge does not confer any separate invariant meaning to $\bar{h}_{tt}$ nor $\bar{h}$.

\section{Linear response}

When submitted to these external forces, the reaction of the black brane (and the dual field theory state) is expected to show up in the holographic stress-energy tensor:  in the tensor channel as an induced shear $T_{ij}$; in the vector channel as a momentum flow $T_{ti}$ due to the dragging by the geometry; and in the scalar channel as local fluctuations in the energy density $T_{tt}$ and averaged pressure $\delta^{ij}T_{ij}$ of the dual plasma.

\subsection{Love numbers}\label{subsec:deflove}

The gauge-invariant content of the response can be readily extracted from the solutions to the master equations using the standard AdS/CFT dictionary. In all three channels, the indices of the differential equation for the variables $Z(v)$ near $v=0$ are $0$ and $n$. Therefore, near the boundary the solutions are expanded as
\beq
Z(v)=A(1+\dots)+B(v^n+\dots)\,.
\eeq
$A$ and $B$ are the coefficients of the non-normalizable and normalizable solutions of the metric perturbation. They depend on $k$, and as is standard in AdS/CFT they correspond, respectively, to the external source acting on the system, and to the expectation value of the operator that the source couples to. In the present case, a non-zero value of $A$ sources a boundary metric deformation $\bar{h}_{\alpha\beta}$ in the corresponding channel, while $B$ determines the response of the system, \ie\ the expectation value of the field theory stress-energy tensor, $\delta T_{\alpha\beta}$, generated by the perturbation.

We define the dimensionless Love numbers $\lambda_{T,V,S}$ for each channel as
\beq\label{deflove}
\lambda=R^n \frac{B}{A}\,.
\eeq
With our normalization \eqref{Zbdry} this is simply $\lambda=B R^n$.

This definition of the Love numbers is in complete analogy to their introduction in the context of asymptotically flat black holes in \cite{Binnington:2009bb}. We can make this more manifest if we change to a radial variable
\beq
r=\frac{R^2}{v},
\eeq
and consider, for instance, a tensor perturbation. Then the corresponding metric component is
\beq
\frac{R^2}{r^2}g_{ij}(r,z)=\delta_{ij}+\bar{h}_{ij}^Te^{ikz}\lp 1+\dots+\lambda_T\frac{R^n}{r^n}+O\lp r^{-n-1}\rp\rp\,,
\eeq
which can be compared to eq.~(1.1) of \cite{Binnington:2009bb}.

\subsection{From Love numbers to stress tensor}

One of the basic entries of the AdS/CFT dictionary (as explained in this context in \cite{Kovtun:2005ev}, see also \cite{Ammon:2015wua}) is that knowledge of the $\lambda$ is tantamount to knowledge of the expectation values of the two-point correlation functions of the stress-energy tensor $T_{\alpha\beta}$. Both are obtained from the terms of order $v^n$ in the series around $v=0$ of the metric coefficients. However the relationship between them is not a simple proportionality. The stress-energy tensor contains contributions besides $\lambda$ that are independent of the boundary condition in the bulk, \ie\ of the specific state of the theory. These contributions are renormalization-scheme dependent. We could, for instance, subtract the vacuum stress-energy out of them, but instead we shall keep these vacuum terms in the counterterm subtraction method. This allows us to retain the effects of vacuum polarization.

Note also that in contrast to the calculation in \cite{Kovtun:2005ev}, which focused on the quasinormal poles of $\langle T_{\alpha\beta}T_{\rho\sigma}\rangle$, we are not setting the source $A$ to zero. Furthermore, we only consider zero-frequency perturbations. Therefore we are investigating properties of the correlation functions $\langle T_{\alpha\beta}T_{\rho\sigma}\rangle$ that do not show up in quasinormal mode analyses.

The correlators $\langle T_{\alpha\beta}T_{\rho\sigma}\rangle$ can be obtained if we know the one-point function $\langle T_{\alpha\beta}\rangle$ as a function of the source, \ie\ of the metric perturbation $\delta\gamma_{\rho\sigma}$, since
\beq
\langle T_{\alpha\beta}T_{\rho\sigma}\rangle=-\frac{2}{\sqrt{-\gamma}}\frac{\delta \langle T_{\alpha\beta}\rangle}{\delta\gamma^{\rho\sigma}}\,.
\eeq

In the gravitational set up $\langle T_{\alpha\beta}\rangle$ is the renormalized holographic stress-energy tensor. For reference, we give its definition in appendix~\ref{app:cterm}. In our case the stress-energy tensor takes the form (henceforth omitting the brackets $\langle\cdots\rangle$)
\beq
T_{\alpha\beta}=T_{\alpha\beta}^0+\delta T_{\alpha\beta}\,,
\eeq
where the first term is the stress-energy tensor of the unperturbed, homogeneous black brane,
\beq
T_{tt}^0=\frac{n-1}{16\pi G}\mu\,,\qquad T_{ij} ^0=\frac{1}{16\pi G}\mu\,\delta_{ij}\,,
\eeq
and the second term $\delta T_{\alpha\beta}$ contains the inhomogeneities linearly induced by the metric deformations $\delta \gamma_{\alpha\beta}=\bar{h}_{\alpha\beta}e^{ik z}$.
Here the bulk Newton constant $G$ is related to the dual theory gauge group's rank $N$ as
\beq
N^2\sim GR^{-3}\quad \textrm{in AdS}_5\,,\qquad N^{3/2}\sim GR^{-2}\quad \textrm{in AdS}_4\,,
\eeq
with numerical factors that depend on the specific realization of the duality (\eg the volume of the compact space transverse to AdS).

Once we compute $\delta T_{\alpha\beta}$ the two-point function can be obtained as
\beq
\langle T_{\alpha\beta}T_{\rho\sigma}\rangle=-2\frac{\partial T_{\alpha\beta}}{\partial\bar{h}^{\rho\sigma}}e^{-ikz}\,.
\eeq

In the following we give the perturbation solutions in a boundary expansion up to order $v^n$, and the form of the stress-energy tensor in terms of $\lambda$. The latter will be computed in later sections.

It is possible to obtain explicit solutions for any $n$, but the expressions are cumbersome so we only give them for AdS$_5$ and AdS$_4$. 

\subsubsection{Boundary expansion and stress-energy tensor in AdS$_5$}\label{subsubsec:bdryads5}

In AdS$_5$ in the tensor sector there are two independent polarizations of the shear, which can be taken to be $h_{\times}=h_{xy}$, and $h_{+}=h_{xx}=-h_{yy}$. For perturbations in the scalar sector we have $h_{xx}=h_{yy}=h/2$. The field theory metric is then
\beq\begin{split}
ds^2=\gamma_{\alpha\beta}dx^\alpha dx^\beta=&~\eta_{\alpha\beta}dx^\alpha dx^\beta
+\bar{h}_{tt}e^{ikz}dt^2+\frac{\bar{h}}2 e^{ikz}(dx^2+dy^2)\\& + 2\bar{h}_{ti}e^{ikz}dt dx^i+\bar{h}_{+}e^{ikz}(dx^2-dy^2)+2\bar{h}_{\times}e^{ikz}dxdy\,.
\end{split}
\eeq

The boundary expansion of $Z$ in the three sectors is the same up to order $v^4$,
\begin{equation}\begin{split}
Z_{T,S,V}(v)=&1-\frac{k^2 v^2}{4}+\left(\frac{\lambda_{T,S,V}}{R^4} -\frac{k^4}{16}  \log v\right) v^4+O\left(v^6\right).
\end{split}\end{equation}

The metric components in the tensor and vector channels are obtained from $Z_{T,V}$ using \eqref{ZTdef} and \eqref{ZVdef}, while for the scalars they are obtained from $Z_S$ and from the solutions of the constraints \eqref{hconst}, \eqref{hzzconst}. We find
\begin{equation}\label{httsol}\begin{split}
h_{tt}(v)=&\bar{h}_{tt}\lp 1+\frac{\mu}2 v^4\rp 
+\frac{\bar{H}}{6}\lp -k^2 v^2 +\left(\frac{4 \lambda_S}{R^4}-4\mu-\frac{k^4}{4}  \log v\right) v^4\rp+O\left(v^6\right),
\end{split}\end{equation}
\begin{equation}\label{hsol}\begin{split}
h(v)=&\bar{h}\lp 1-\frac{\mu}2 v^4\rp
+\frac{\bar{H}}{6}\lp -k^2 v^2 +\left(\frac{4 \lambda_S}{R^4}+2\mu-\frac{k^4 }{4} \log v\right) v^4\rp+O\left(v^6\right),
\end{split}\end{equation}
\begin{equation}\begin{split}
h_{zz}(v)=&\bar{h}_{zz}+\lp \frac{\bar{h}_{tt}}{2}-\frac{\bar{H}}{3}\rp k^2 v^2+\frac{\bar{h}_{tt}}2\,\mu  v^4+O\left(v^6\right).
\end{split}\end{equation}

The stress-energy tensor is
\beq\label{Tmunu5}\begin{split}
8\pi G\, T_{\alpha\beta}\,dx^\alpha dx^\beta=&\lp 3 dt^2+dx^2+dy^2+dz^2\rp \frac{\mu}2\lp  1+\bar{h}_{tt}e^{i k z} \rp\\
&+\lp 2\bar{h}_{\times}\,dx\, dy+\bar{h}_{+}(dx^2-dy^2)\rp e^{ikz}\lp \frac{2\lambda_T}{R^4}+\frac{\mu}2-\frac{3k^4}{32}\rp\\
&+ 2\bar{h}_{ti}\,dt\, dx^i\,e^{ikz}\, \left(\frac{2\lambda_V}{R^4}+\frac{\mu}{2}-\frac{3k^4}{32}  \right)\\
&+ \bar{H} dt^2 e^{ikz} \lp \frac43 \lp\frac{\lambda_S}{R^4}-\mu\rp-\frac{k^4}{16} \rp\\
&+\bar{H} \lp dx^2+dy^2\rp e^{ikz} \lp \frac{2\lambda_S}{3R^4}-\frac{\mu}6-\frac{k^4}{32} \rp\\
&+ dz^2\,\frac{\mu}{2}\bar{h}_{tt} e^{i k z} \,.
\end{split}\eeq
The $k^4$ terms here are renormalization-scheme dependent, and in general are modified to $k^4\to k^4 (1-4b/3)$, where the arbitrary constant $b$ is the coefficient of the finite counterterms in \eqref{holotab}. In the following we fix $b=0$ for simplicity, but the existence of this ambiguity should be borne in mind.

The gauge-invariant boundary scalar is
\beq\label{defH}
\bar{H}=\bar{h}_{tt}+\frac{\bar{h}}2\,.
\eeq
As we discussed in the previous section, in the scalar sector only this parameter is physically meaningful, while $\bar{h}_{tt}$ and $\bar{h}$ separately are not: the coordinate transformations \eqref{vchange} change them. Consistently with this, observe that if we rescale
\beq\label{muchange}
\mu \to \mu\lp 1- \bar{h}_{tt} e^{ikz}\rp\,,
\eeq
and also perform a rescaling of $z$ (which makes $\bar{h}_{zz}\neq 0$), then we can make $\bar{h}_{tt}$ disappear from \eqref{Tmunu5}. In other words, the apparent spatial dependence of the plasma temperature does not have any invariant meaning for a CFT.
Even if \eqref{muchange} suggests that the perturbation makes the horizon position $z$-dependent, this is a gauge effect. In particular it is easy to see that the surface gravity remains uniform over the horizon, as required by the zeroth law. 

We can also write the stress-energy tensor in a way that separates its different contributions and connects more directly to the hydrodynamic expansion at small $k$. Define a boundary velocity field $u^\alpha$ as
\beq\label{uLandau}
u^t=1+\frac{e^{ikz}}2 \bar{h}_{tt},\qquad u^i=-\lp \frac{\lambda_V}{R^4\mu}+1-\frac{3k^4}{64\mu}\rp e^{ikz}\bar{h}_{ti}\,,
\eeq
which is unit-normalized, $\gamma^{\alpha\beta}u_\alpha u_\beta=-1$,
and choose
\beq\label{landauhtt}
\bar{h}_{tt}=-\bar{H}\lp\frac{4}{9}\lp \frac{\lambda_S}{R^4\mu}-1\rp-\frac{k^4}{48\mu}\rp\,.
\eeq
Then the stress-energy tensor takes a `Landau frame' form
\beq\label{landaustress}
T_{\alpha\beta}=\frac{\mu}{16\pi G}\lp \gamma_{\alpha\beta}+4u_\alpha u_\beta\rp+
T_{\alpha\beta}^{(1)}\,,
\eeq
in which the first term has the form of a perfect-fluid stress-energy tensor (with conformal equation of state) and the second term is purely spatial, orthogonal to $u^\alpha$,
\beq
u^\alpha T_{\alpha\beta}^{(1)}=0\,.
\eeq
It is given by
\beq\label{landauT1}\begin{split}
8\pi G\, T_{\alpha\beta}^{(1)}\,dx^\alpha dx^\beta=
&\lp 2\bar{h}_{\times}\,dx\, dy+\bar{h}_{+}(dx^2-dy^2)\rp e^{ikz}\lp \frac{2\lambda_T}{R^4}-\frac{3k^4}{32}\rp\\
&+\bar{H} \lp dx^2+dy^2-2dz^2\rp e^{ikz} \lp \frac{2}{9}\lp \frac{\lambda_S}{R^4}-\mu\rp-\frac{k^4}{96} \rp \,.
\end{split}
\eeq

When the stress-energy tensor is written in this way, the first part can be regarded as capturing how the plasma adapts to the deformed geometry $\gamma_{\alpha\beta}$ and to a velocity flow $u_\alpha$ while maintaining its perfect-fluid form. The choice of $u$ and of $\bar{h}_{tt}$ is indeed such that the vector-channel polarization, and the scalar-channel polarization in the $tt$ direction, are all encoded in this term. 
The second term, $T_{\alpha\beta}^{(1)}$, measures the polarization effects away from the perfect-fluid form. Bear in mind, though, that both terms in \eqref{landaustress} contain physical polarizations of the uniform plasma.

We will see that when $k\to 0$ we have
\beq\label{offsets5}
\lambda_T\to 0,\qquad \lambda_V\to -\mu R^4,\qquad \lambda_S\to\mu R^4\,. 
\eeq 
This implies that in the limit that the perturbation is homogeneous we have $u^\alpha\to \delta^\alpha{}_t$ and $T_{\mu\nu}^{(1)}\to 0$, and hence there does not remain any physical polarization effect.

\subsubsection{Boundary expansion and stress-energy tensor in AdS$_4$}

In AdS$_4$ there are no tensor perturbations. In the scalar sector,
$h(v)=h_{xx}(v)$. The field theory metric is 
\beq\begin{split}
ds^2=\gamma_{\alpha\beta}dx^\alpha dx^\beta=\eta_{\alpha\beta}dx^\alpha dx^\beta
+\bar{h}_{tt}e^{ikz}dt^2+2\bar{h}_{tx}e^{ikz}dt dx+\bar{h}_{xx}e^{ikz} dx^2\,.
\end{split}
\eeq

The boundary expansion for $Z$ is
\begin{equation}\begin{split}
Z_{V,S}(v)=&1-\frac{k^2 v^2}{2}+\frac{\lambda_{V,S}}{R^3}  v^3+O\left(v^4\right),
\end{split}\end{equation}
and the stress tensor
\beq\label{Tmunu4}\begin{split}
8\pi G\, T_{\alpha\beta}dx^\alpha dx^\beta=&\lp 2 dt^2+dx^2+dz^2\rp \frac{\mu}2\lp  1+\frac{\bar{h}_{tt}}2e^{i k z} \rp\\
&+ 2\bar{h}_{ti}\,dt\, dx^i\,e^{ikz}\frac32\lp\frac{\lambda_V}{R^3}+\frac{\mu}3\rp \\
&+ \bar{H} dt^2 e^{ikz} \frac34\lp\frac{\lambda_S}{R^3}-\frac{\mu}2\rp \\
&+\bar{H} dx^2 e^{ikz} \frac34\lp\frac{\lambda_S}{R^3}+\frac{\mu}6\rp \\
&+ dz^2\,\frac{\mu}{2}\bar{h}_{tt} e^{i k z} \,.
\end{split}\eeq

Now the gauge-invariant boundary scalar is
\beq
\bar{H}=\bar{h}_{tt}+\bar{h}_{xx}\,,
\eeq
and the metric functions are
\begin{equation}\begin{split}
h_{xx}(v)=&\bar{h}_{xx}\lp 1-\frac{\mu}2 v^3\rp+\frac{\bar{H}}{4}\lp -k^2 v^2+\lp \frac{2\lambda_S}{R^3} +\mu\rp v^3\rp+O\left(v^4\right),
\end{split}\end{equation}
\begin{equation}\begin{split}
h_{tt}(v)=&\bar{h}_{tt}+\frac{\bar{H}}{4}\lp -k^2 v^2+\lp \frac{2\lambda_S}{R^3} -\mu\rp v^3\rp+O\left(v^4\right),
\end{split}\end{equation}
\begin{equation}\begin{split}
h_{zz}(v)=&\bar{h}_{zz}+\lp \frac{\bar{h}_{tt}}{2}-\frac{\bar{H}}{4}\rp k^2 v^2+\frac{\bar{h}_{tt}}2\,\mu  v^3+O\left(v^4\right).
\end{split}\end{equation}

Similar remarks as in AdS$_5$ apply about the elimination of $\bar{h}_{tt}$.

The `Landau frame' expression of the stress-energy tensor is
\beq\label{landaustress4}
T_{\alpha\beta}=\frac{\mu}{16\pi G}\lp \gamma_{\alpha\beta}+3u_\alpha u_\beta\rp+
T_{\alpha\beta}^{(1)}\,,
\eeq
with
\beq\label{uLandau4}
u^t=1+\frac{e^{ikz}}2 \bar{h}_{tt},\qquad u^i=-\lp \frac{\lambda_V}{R^3\mu}+1\rp e^{ikz}\bar{h}_{ti}\,,
\eeq
\beq\label{landauhtt4}
\bar{h}_{tt}=-\frac{\bar{H}}{2}\lp\frac{\lambda_S}{R^3\mu}-\frac12\rp\,,
\eeq
and
\beq\label{landauT14}
8\pi G\, T_{\alpha\beta}^{(1)}\,dx^\alpha dx^\beta=
\frac{3\bar{H}}{8} (dx^2-dz^2) e^{ikz} \lp\frac{\lambda_S}{R^3}-\frac{\mu}2\rp  \,.
\eeq
Again, when $k\to 0$ we will find
\beq\label{offsets4}
\lambda_V\to -\mu R^3,\qquad \lambda_S\to\frac{\mu R^3}{2}\,, 
\eeq 
which cancel the zero-momentum offsets in $u^\alpha$ and $T_{\mu\nu}^{(1)}$.

\section{Vacuum polarization}

Let us now turn to the explicit calculation of the Love numbers. 

It is instructive to begin with the polarization of the vacuum, since it can be solved exactly in all channels, for all $k$, and in all dimensions. These Love numbers can be regarded as representing Casimir-like stress-energies of the field theory vacuum.

In the vacuum state, with $\mu=0$, the equations in the three channels become the same,
\begin{equation}
Z''(v)-\frac{n-1}{v}Z'(v)-k^2 Z(v)=0.
\end{equation}
This equation is solved in terms of modified Bessel functions. The solution that remains finite at the Poincar\'e horizon, $v\to \infty$, is
\beq
Z(v) = v^{n/2} K_{n/2}(k v)\,.
\eeq
Expanding this solution in series around $v=0$ we obtain the vacuum Love numbers,
\begin{equation}\label{lovevac}
\lambda_\textrm{vac}(k) = 
\begin{cases}
\left(H_{n/2}-2 \gamma -2\log \left(\frac{kR}{2}\right)\right) \dfrac{(-1)^{n/2} }{(n/2-1)! \, (n/2)! \, 2^n} \, (kR)^n  \quad &n \text{ even}\\
\dfrac{\Gamma (-n/2)}{\Gamma (n/2)\, 2^{n}}\, (kR)^n \quad &n \text{ odd}
\end{cases}
\end{equation}
where $\gamma$ is the Euler-Mascheroni constant and $H_n=\sum_{p=1}^n p^{-1}$ are the harmonic numbers. 

Observe that: (i) the dependence $\sim (kR)^n$ is the one expected for the vacuum energy density of a conformal field theory in $n$ dimensions; (ii) the logarithmic term in even $n$ comes from the conformal anomaly and makes the terms $H_{n/2}-2 \gamma$ scheme dependent; (iii) the sign of the Love numbers (at large enough $k$) alternates as $n\to n+2$. This dimension-dependence of the sign of the polarization response is the same as for the Casimir energy on a spherical space \cite{Emparan:1999pm}.

In the specific cases of interest to us here,
\begin{eqnarray}
\lambda_\textrm{vac}(k) &=& \frac{(kR)^3}{3} \quad \text{in AdS}_4\,,\label{vac4}\\
\lambda_\textrm{vac}(k) &=&-\frac{(kR)^4}{16}\lp \log\left(\frac{kR}{2}\right)+\gamma-\frac34 \rp  \quad \text{in AdS}_5\,.\label{vac5}
\end{eqnarray}

For large $k$ the perturbations probe the ultraviolet, short-distance structure of the field theory and the results should be asymptotically independent of whether the state is at finite or zero temperature. In other words, for $k\gg T$ the perturbations concentrate in the bulk around $0\leq v\lesssim 1/k$ and are largely insensitive to the presence or absence of the brane. It then follows that the Love numbers at large $k$ should always asymptote to their conformal vacuum values, and in particular 
\beq\label{largek}
\lambda(k)\sim (-1)^{\lfloor n/2\rfloor+1}(kR)^n\,.
\eeq

Finally, note that when $\mu=0$ the gauge transformations \eqref{vchange} do not introduce any non-analytic behavior in the bulk. The gauge is analytic for any arbitrary choice of $\bar{h}_{tt}$.

\section{Polarization of the finite-temperature plasma}\label{sec:finT}

At finite temperature the perturbation equations do not admit exact solutions. We solve them in two ways: in a long-wavelength, hydrodynamic expansion for small $k$, and numerically for a range of $k$, up until the large-$k$ asymptotic behavior \eqref{largek} is established.

\subsection{Long-wavelength expansion}

The solution is obtained by a conventional perturbative expansion in powers of $k$. The results for the metric functions are given in appendix~\ref{app:hydro}.

The Love numbers that we find are
\paragraph{AdS$_5$:}
\begin{equation}\label{LaT5}
\frac{\lambda_T(k)}{R^4} = \frac{k^2\sqrt{\mu}}{8}+\frac{k^4}{64} \lp 3-4 \log 2\rp-\frac{k^6}{768\sqrt{\mu}}  \left(\pi ^2-12 \lp\log 2\rp^2\right) + O(k^8),
\end{equation}
\begin{equation}\label{LaV5}
\frac{\lambda_V(k)}{R^4} = -\mu+\frac{k^2\sqrt{\mu}}{4}-\frac{k^4}{64}+\frac{k^6}{128\sqrt{\mu}}  \lp1-2\log 2\rp -\frac{k^8}{6144\mu}\left(\pi ^2+6-24 \log 2\right)  +O(k^{10}),
\end{equation}
\begin{equation}\label{LaS5}
\frac{\lambda_S(k)}{R^4} = \mu-\frac{3 k^2\sqrt{\mu}}{8}+\frac{k^4}{64}\lp11-4\log 2\rp + O(k^6).
\end{equation}

\paragraph{AdS$_4$:}
\begin{equation}\label{LaV4}
\frac{\lambda_V(k)}{R^3} = -\mu+\frac{k^2\mu^{1/3}}{2}+\frac{k^4}{12\mu^{1/3}}  +\frac{k^6}{72\mu}  \left(\sqrt{3} \pi -9+3\log 3\right) + O(k^8),
\end{equation}
\begin{equation}\label{LaS4}
\frac{\lambda_S(k)}{R^3} = \frac{\mu}{2}+\frac{2 k^4}{9\mu^{1/3}}-\frac{k^6}{27\mu} + O(k^8).
\end{equation}
\medskip

Some comments are in order. First, observe that since this is a small $k$ expansion in $k/T\sim k /\mu^{1/n} \ll 1$, we do not expect to recover the large-$k$ asymptotic behavior \eqref{largek} of the vacuum.

Second, as anticipated in \eqref{offsets5} and \eqref{offsets4}, we find non-zero values of the vector and scalar Love numbers at very long wavelengths, $k\to 0$. These are such that the physical polarization effects vanish in this limit.

Finally, let us compare these results with those in \cite{Bhattacharyya:2008ji,Bhattacharyya:2008mz} for the gravitational forcing on the AdS black brane in the hydrodynamic limit. Refs.~\cite{Bhattacharyya:2008ji,Bhattacharyya:2008mz} give
\beq\label{fluidTmunu}
T_{\alpha\beta}=\frac{\mu}{16\pi G}\lp \gamma_{\alpha\beta}+n u_\alpha u_\beta\rp+
\frac{\mu^{\frac{n-2}{n}}}{8\pi G} C_{\alpha\gamma\beta\delta}u^\gamma u^\delta\,.
\eeq
Here $C_{\alpha\mu\beta\nu}$ is the Weyl tensor of the field theory metric $\gamma_{\alpha\beta}$, and the velocity vector $u^\alpha$ is chosen in the Landau frame. This result is valid to two-derivative order in the boundary theory, hence to order $k^2$ in the linearized approximation. It is straightforward to compare the Weyl term against our result \eqref{landauT1} up to this order, and verify the agreement between the two calculations in AdS$_5$.
In AdS$_4$ the boundary Weyl tensor is identically zero, so $T_{\alpha\beta}^{(1)}$ vanishes at order $k^2$. This is in agreement with the absence of a $k^2$ term in $\lambda_S$ in \eqref{LaS4}.\footnote{Refs.~\cite{Bhattacharyya:2008ji,Bhattacharyya:2008mz} work in Eddington-Finkelstein coordinates which are regular at the horizon. In our calculations, in AdS$_5$ the analytic gauge choice \eqref{htt5} coincides up to order $k^2$ with the Landau gauge \eqref{landauhtt}.
In AdS$_4$ the Landau gauge \eqref{landauhtt4} does not coincide with the analytic gauge \eqref{htt4} at order $k^2$. However, it seems that this could be remedied if in \eqref{landaustress4} we redefined $\mu\to \mu(1+ c\, e^{ikz})$ with suitably chosen $c=O(\bar{h}_{\alpha\beta},k^2)$.}

\subsection{Numerical results}

Now we solve the equations by numerical integration. After setting, without loss of generality, $\mu=1$, we impose regularity on the horizon at $v=1$ by demanding that the gauge invariant function $Z(v)$ is analytic there.
Then we solve the equations in powers of $(1-v)$ to a high order (without any arbitrary constants other than the overall normalization of $Z$), and proceed to integrate them numerically towards the boundary, where we extract the Love numbers \eqref{deflove}. We do the integrations with the \texttt{NDSolve} function from \textsc{Mathematica}, which uses a fourth-order Runge-Kutta procedure with adaptive step. The equations are very well behaved so the calculation is unproblematic.

\begin{figure}[t]
\begin{center}
\includegraphics[width=200pt]{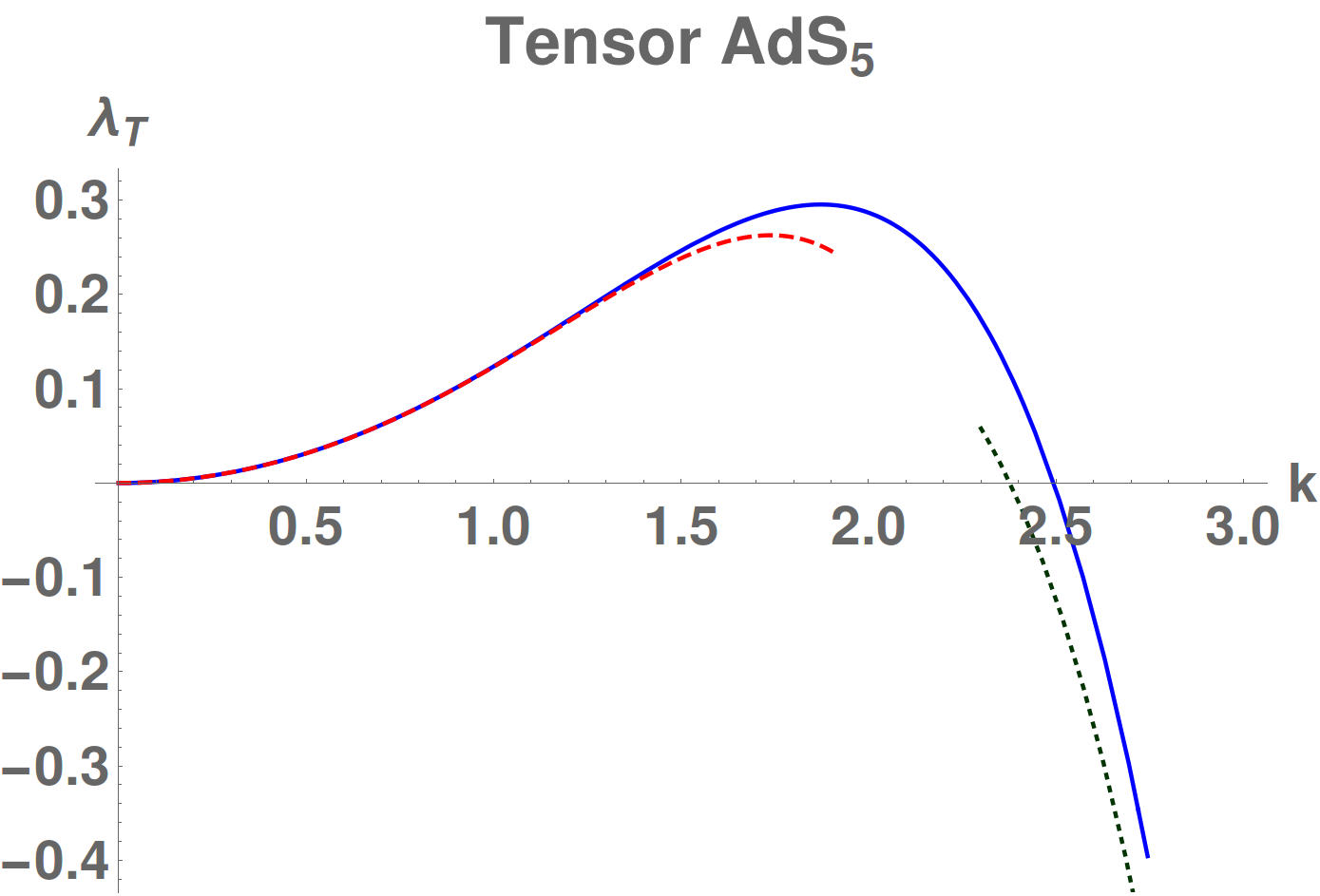}\qquad
\includegraphics[width=200pt]{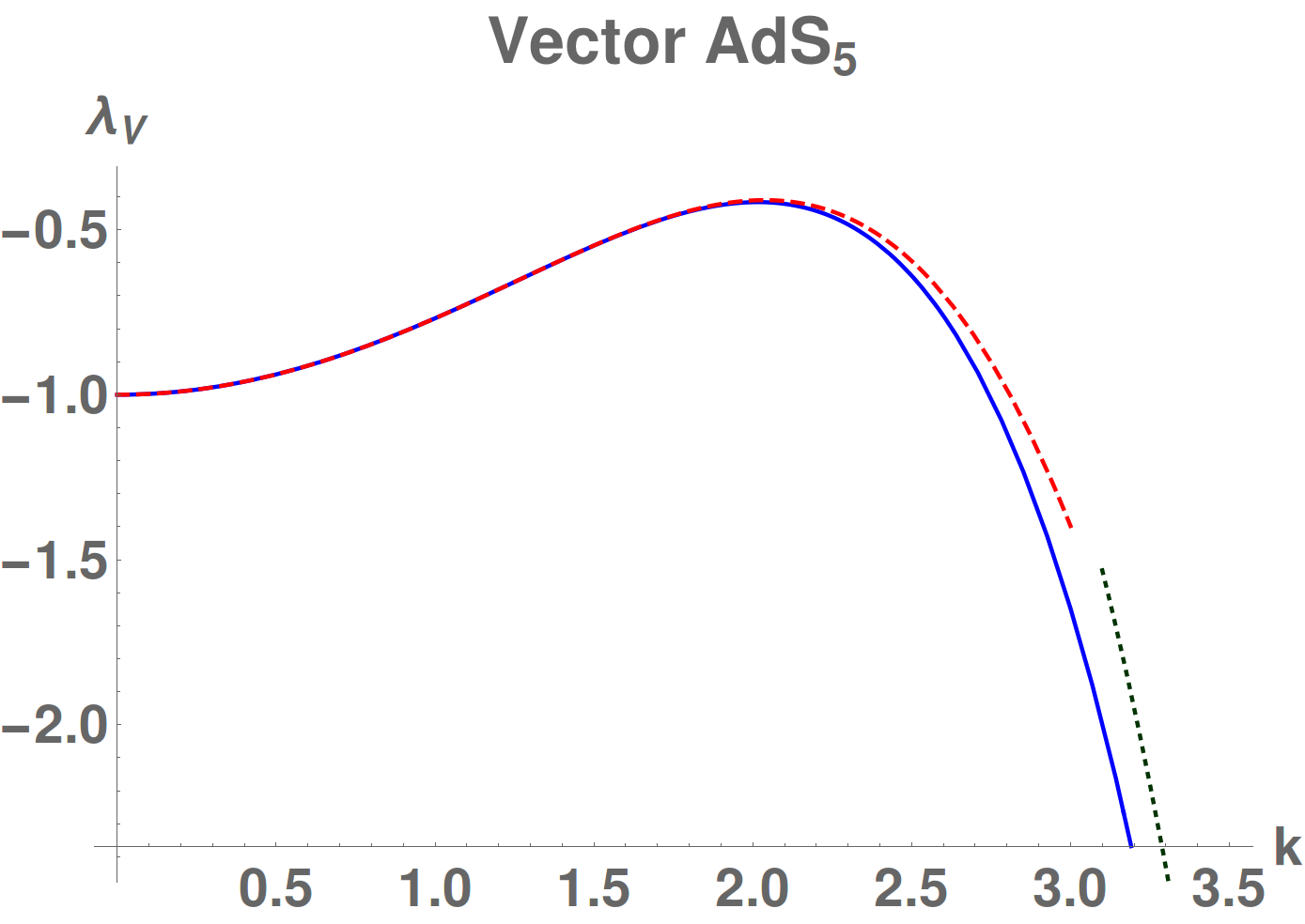}\\
\bigskip
\includegraphics[width=200pt]{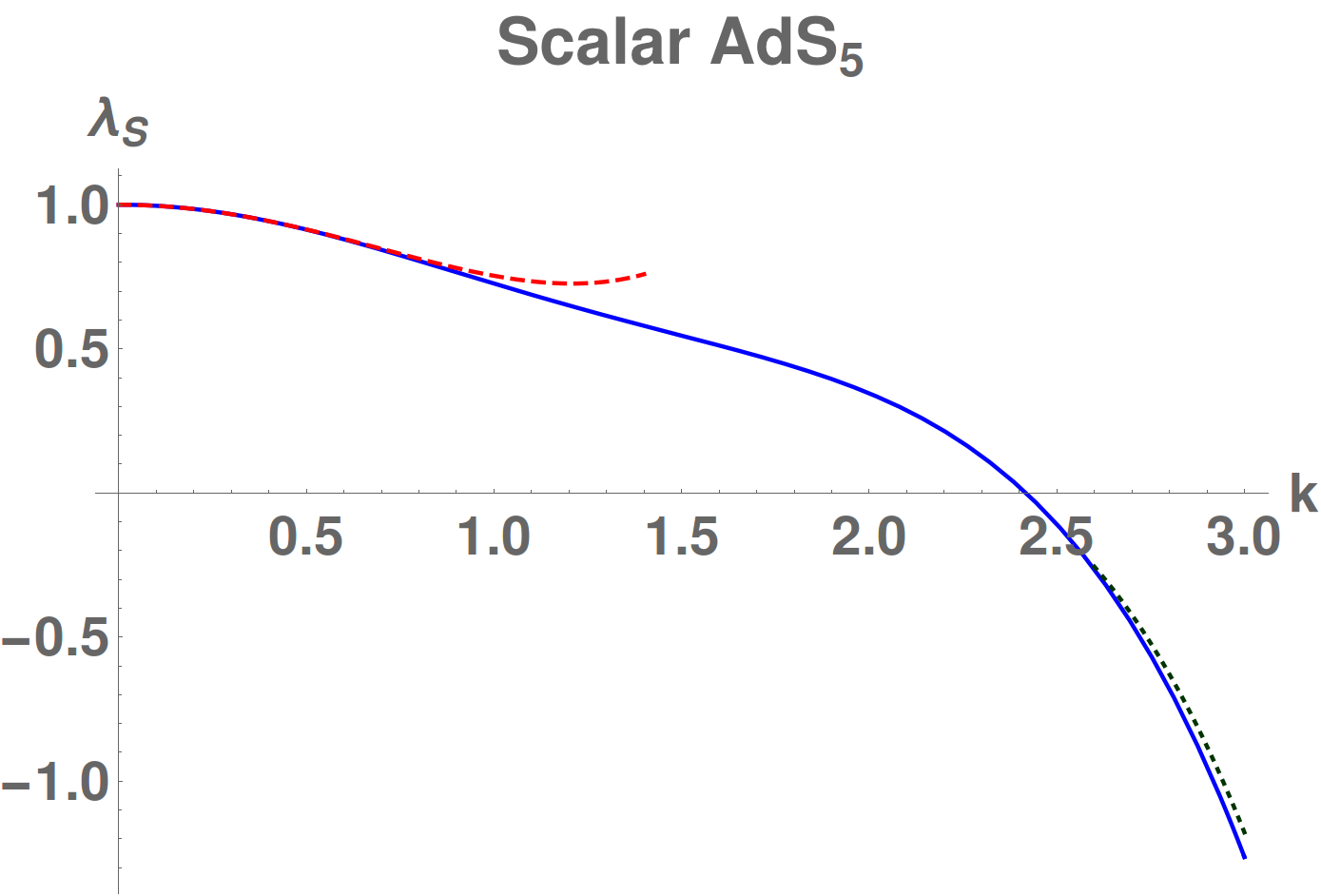}
\caption{\small Love numbers $\lambda_{T,V,S}$ for black branes in AdS$_5$ as a function of the wavenumber $k$. Solid blue: numerical results. Dashed red: perturbative expansions in powers of $k$, eqs.~\eqref{LaT5}, \eqref{LaV5}, \eqref{LaS5}. Dotted green: large-$k$ limit \eqref{vac5}. We set $R=1$, the Love numbers $\lambda_{T,V,S}$ are dimensionless, and $k$ is measured in units of $\mu^{1/4}=\pi T$. \label{fig:AdS5LovePlots}}
\end{center}
\end{figure}

\begin{figure}[t]
\begin{center}
\includegraphics[width=200pt]{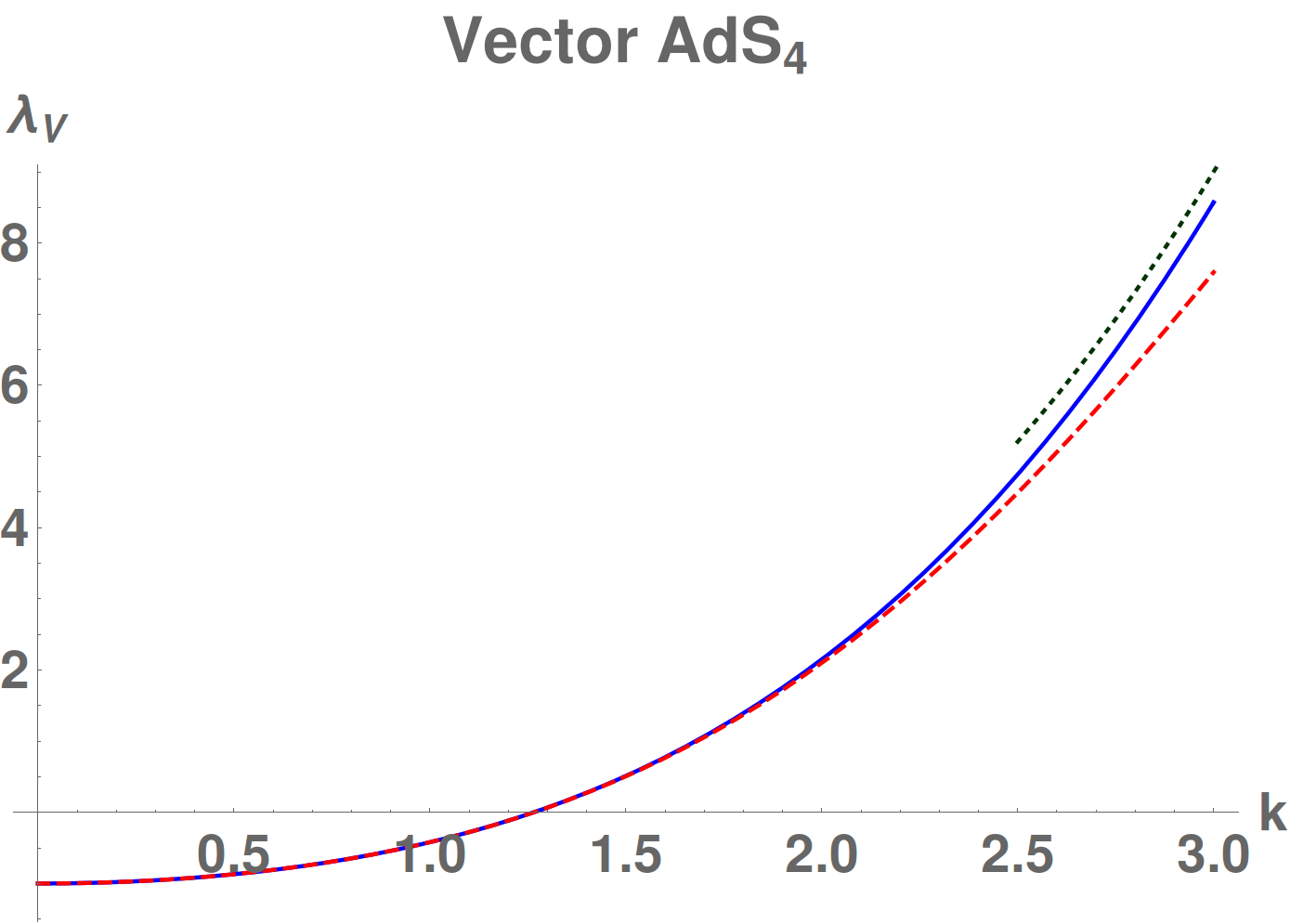}\qquad
\includegraphics[width=200pt]{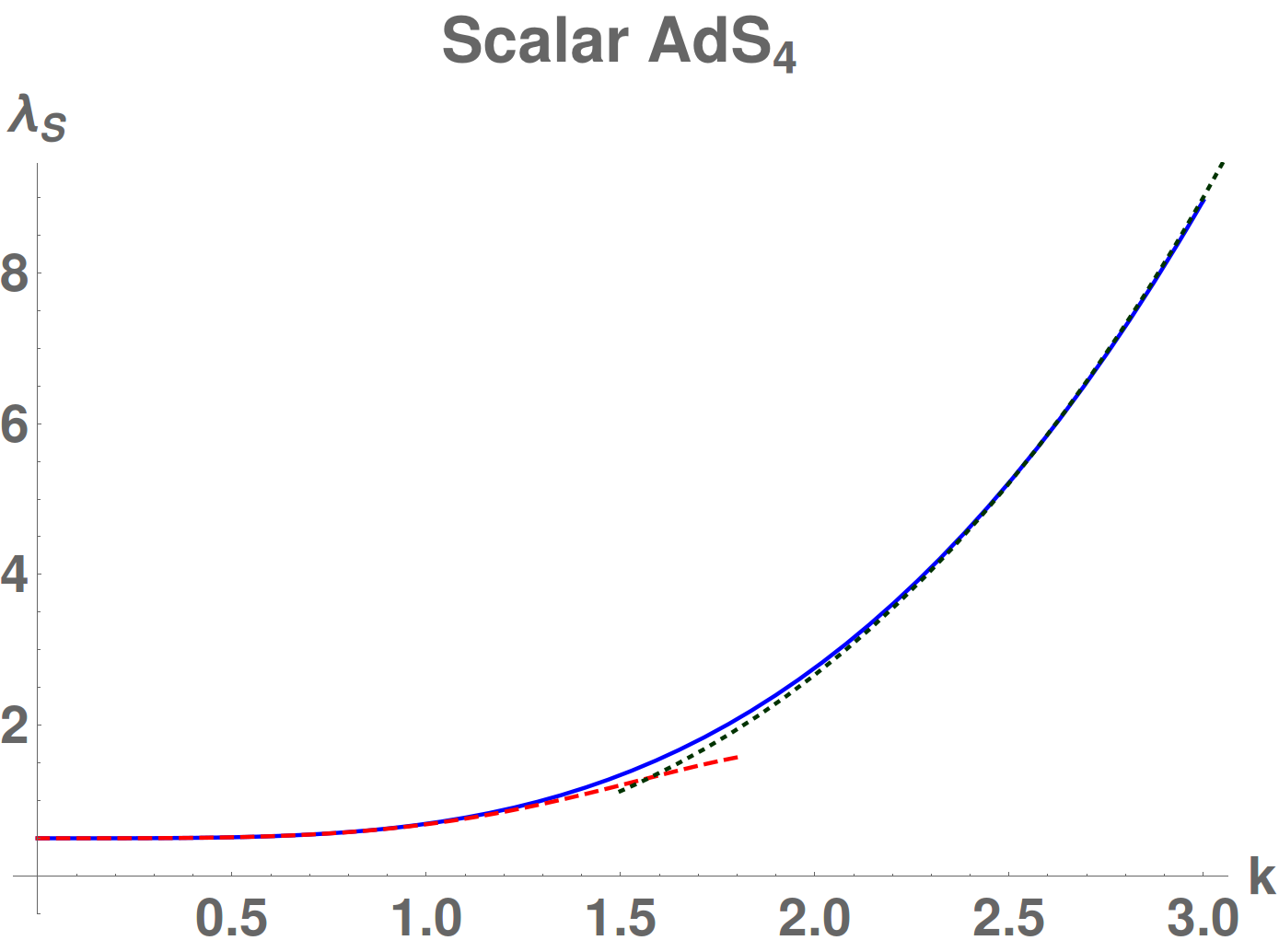}
\caption{\small Love numbers $\lambda_{V,S}$ for black branes in AdS$_4$ as a function of the wavenumber $k$, in units $\mu=1$.  Solid blue: numerical results. Dashed red: perturbative expansions in powers of $k$, eqs.~\eqref{LaV4}, \eqref{LaS4}. Dotted green: large-$k$ limit \eqref{vac4}. The Love numbers $\lambda_{V,S}$ are dimensionless and $k$ is measured in units of $\mu^{1/3}=4\pi T/3$. \label{fig:AdS4LovePlots}}
\end{center}
\end{figure}

The results are shown in figs.~\ref{fig:AdS5LovePlots}, \ref{fig:AdS4LovePlots}, where we compare them with the hydrodynamic expansion at small $k$ and with the large-$k$ vacuum limit. In appendix~\ref{app:angauge} we give the values of $\bar{h}_{tt}(k)$ that result when we choose a gauge in which $h_{tt}(v)$ (and then also $h(v)$ and $h_{zz}(v)$) is analytic at the horizon.

Overall, we see that the small-$k$ hydrodynamic expansion and the large-$k$ values from the vacuum provide together a good approximation to the numerical calculations. It seems likely that Pad\'e approximants can interpolate efficiently at intermediate values of $k$, but we have not attempted this.

Observe that the Love numbers can change sign as $k$ increases, \ie\ the plasma appears to polarize in opposite ways at small and large wavelengths. This must be interpreted with care, given that the zero-momentum offsets in $\lambda$, \eqref{offsets5} and \eqref{offsets4}, disappear in the stress-energy tensor in Landau frame. The latter may be more appropriate to study the sign of the response. Then we see, for instance, that the anisotropic, transverse pressure induced in the scalar channel, $T^{(1)}_{xx}+T^{(1)}_{yy}$, is negative for all $k$ in AdS$_5$, and positive for all $k$ in AdS$_4$. The (gauge-dependent) term $\bar{h}_{tt}$ which, in Landau frame, reflects the perfect-fluid response in the scalar sector, has opposite signs in AdS$_5$ and AdS$_4$, but in each case it retains the same sign for all $k$.
On the other hand, the vector-channel velocity $u^i$ induced in AdS$_5$ changes sign as $k$ is increased, while in AdS$_4$ it keeps the same orientation at all $k$. 

Perhaps the most salient feature is that the response coefficients in AdS$_4$ show a mostly featureless monotonicity in $k$, while in AdS$_5$ the behavior differs significantly at large and small $k$. This occurs even for the vacuum polarization, \eqref{vac5}, but in this case it is the $\log k$ in the Love number, and not a power of $k$, that effects the change. 

As is familiar from the Casimir effect, the sign of quantum polarization effects is often difficult to anticipate on intuitive grounds. Nevertheless, it may be interesting to investigate further the possible meaning of these results. The exploration of further models might hint at universal features of the geometric polarization. 

\section{Electric polarization}

Now we consider the polarizing effect on the black brane of a small static electric field in the $z$ direction, with electric potential $A_t(v)e^{ikz}$. The dual plasma, initially neutral, polarizes into an inhomogeneous distribution of positive and negative charge densities due to the presence of an external chemical potential. We denote the amplitude of the chemical potential by
\beq
\bar{A_t}=A_t(0)\,,
\eeq
and, like in our previous analysis, we introduce the variable $Z_E$ by
\beq
A_t(v)=\bar{A_t}\, Z_E(v)\,.
\eeq

\subsection{Linear response theory}

The Maxwell equations in the black brane background are
\begin{equation}\label{maxwelleq}
Z_E''(v)-\frac{n-3}{v}Z_E'(v)-\frac{k^2}{f} Z_E(v)=0\,.
\end{equation}
The boundary expansion of the solutions takes the form
\beq
Z_E(v)=A(1+\dots)+B(v^{n-2}+\dots)\,,
\eeq
and the polarization response is determined by the coefficient
\beq
\lambda_E=R^{n-2}\frac{B}{A}\,.
\eeq
This coefficient determines the expectation value of the charge density $J^t$. In order to find the precise relation, following the standard AdS/CFT prescription we differentiate the Maxwell action with respect to the boundary electric potential to get
\beq
\langle J^t\rangle =-\frac12 \sqrt{-\hat{g}}\,n_\mu F^{\mu t}\,,
\eeq
where $n_\mu$ is the unit normal to the boundary at small $v$ with induced metric $\hat{g}_{\alpha\beta}$. The charge density at the boundary is then given by the electric field in the normal direction.

In AdS$_4$ the boundary expansion of the solution to \eqref{maxwelleq} is
\beq
Z_E(v)=1+\frac{\lambda_E}{R} v +O(v^2)
\eeq
which yields
\beq
\langle J^t\rangle = \bar{A_t}e^{ikz}\frac{\lambda_E}{2R}\,.
\eeq

In AdS$_5$ there is a logarithmic term
\beq
Z_E(v)=1+\lp \frac{\lambda_E}{R^2} +\frac{k^2}2\log v\rp v^2 +O(v^3)\,.
\eeq
This results in a divergence that is cancelled by adding a boundary counterterm to the action of the form $I_{ct}\sim \log v\,\int d^4x \sqrt{-\hat{g}} F_{\alpha\beta} F^{\alpha\beta}$. Then
\beq
\langle J^t\rangle =\bar{A_t}e^{ikz}\lp \frac{\lambda_E}{R^2} +\frac{k^2}4\rp
\eeq
(again, the term $k^2$ is renormalization-scheme dependent).

The two-point correlation function is obtained as
\beq
\langle J^t J^t\rangle=\frac{\delta\langle J^t\rangle}{\delta \bar{A_t}}e^{-ikz}\,.
\eeq

\subsection{Polarization coefficients}

In the zero-temperature vacuum, $\mu = 0$, eq.~\eqref{maxwelleq} becomes
\begin{equation}
Z_E''(v)-\frac{n-3}{v}Z_E'(v)-k^2 Z_E(v)=0,
\end{equation}
which is the same as the one for gravitational perturbations if we change $n \to n-2$. Therefore, the electric polarization of the vacuum can be determined from the gravitational vacuum Love numbers as
\beq
\lambda_{E,\textrm{vac}}^{(n)}(k) = \lambda_{\textrm{vac}}^{(n-2)}(k)
\eeq
and the latter were computed in \eqref{lovevac}. This gives
\begin{equation}\label{vacel}\begin{split}
\lambda_{E,\textrm{vac}} =&  -kR\quad \text{in AdS}_4\,,\\ \lambda_{E,\textrm{vac}} =&-\frac{(kR)^2}{2}\lp \log\left(\frac{kR}{2}\right)+\gamma-\frac12 \rp \quad \text{in AdS}_5\,.
\end{split}
\end{equation}

At finite temperature, the long-wavelength hydrodynamic expansion yields \begin{equation}\label{LaE5}
\frac{\lambda_E(k)}{R^2} = -\sqrt{\mu}+\frac{k^2}{4}  \lp 2\log 2-1\rp   +\frac{k^4}{96\sqrt{\mu}}  \left(\pi ^2-12 ( \log 2)^2\right)   +O(k^6) \quad \textrm{in AdS}_5\,,
\end{equation}
and
\begin{equation}\label{LaE4}
\frac{\lambda_E(k)}{R} = -\mu^{1/3} +\frac{k^2}{6\mu^{1/3}}\lp 3\log 3-\sqrt{3}\pi\rp + O(k^4) \quad \textrm{in AdS}_4\,.
\end{equation}

\begin{figure}[t]
\begin{center}
\includegraphics[width=200pt]{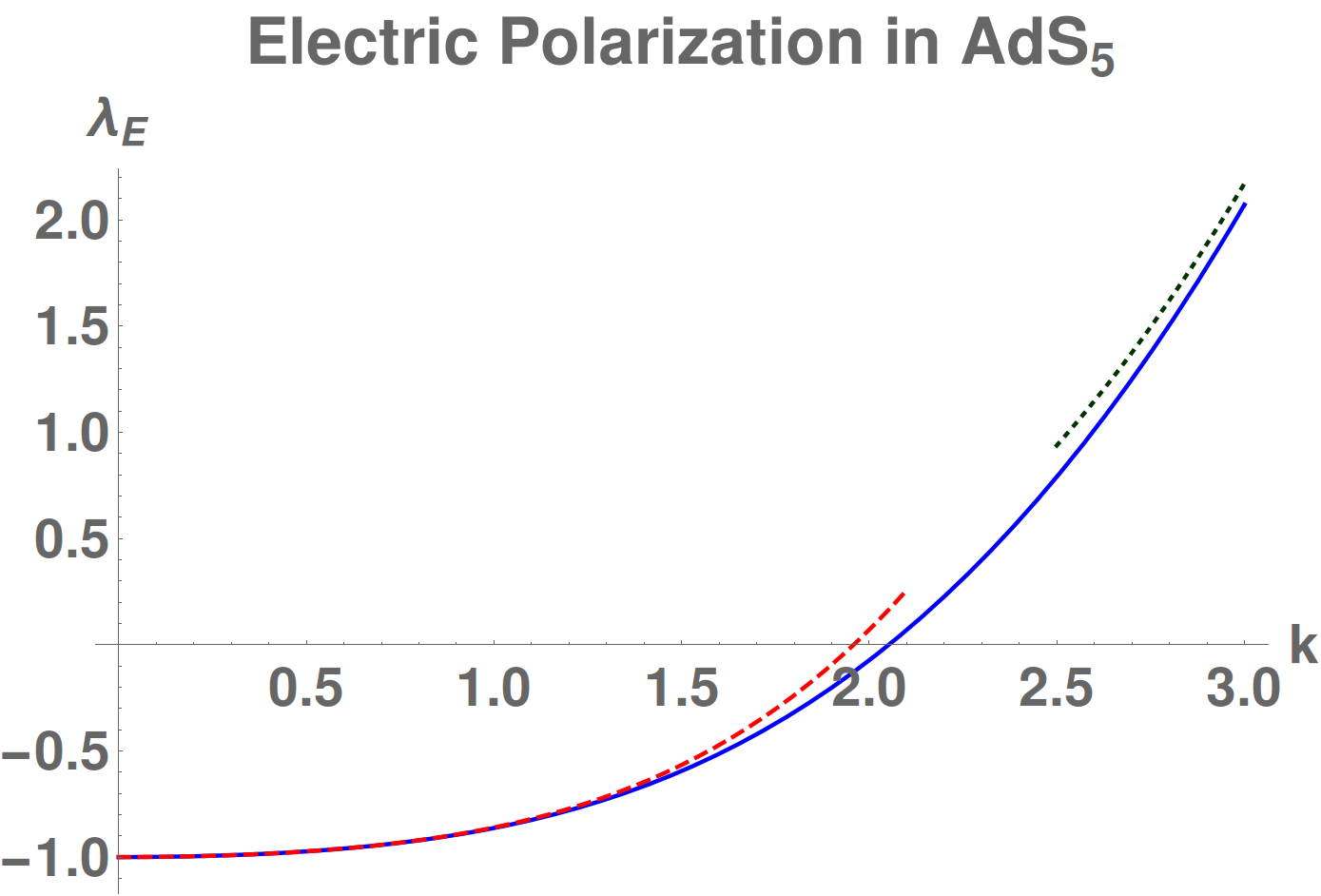}\qquad
\includegraphics[width=200pt]{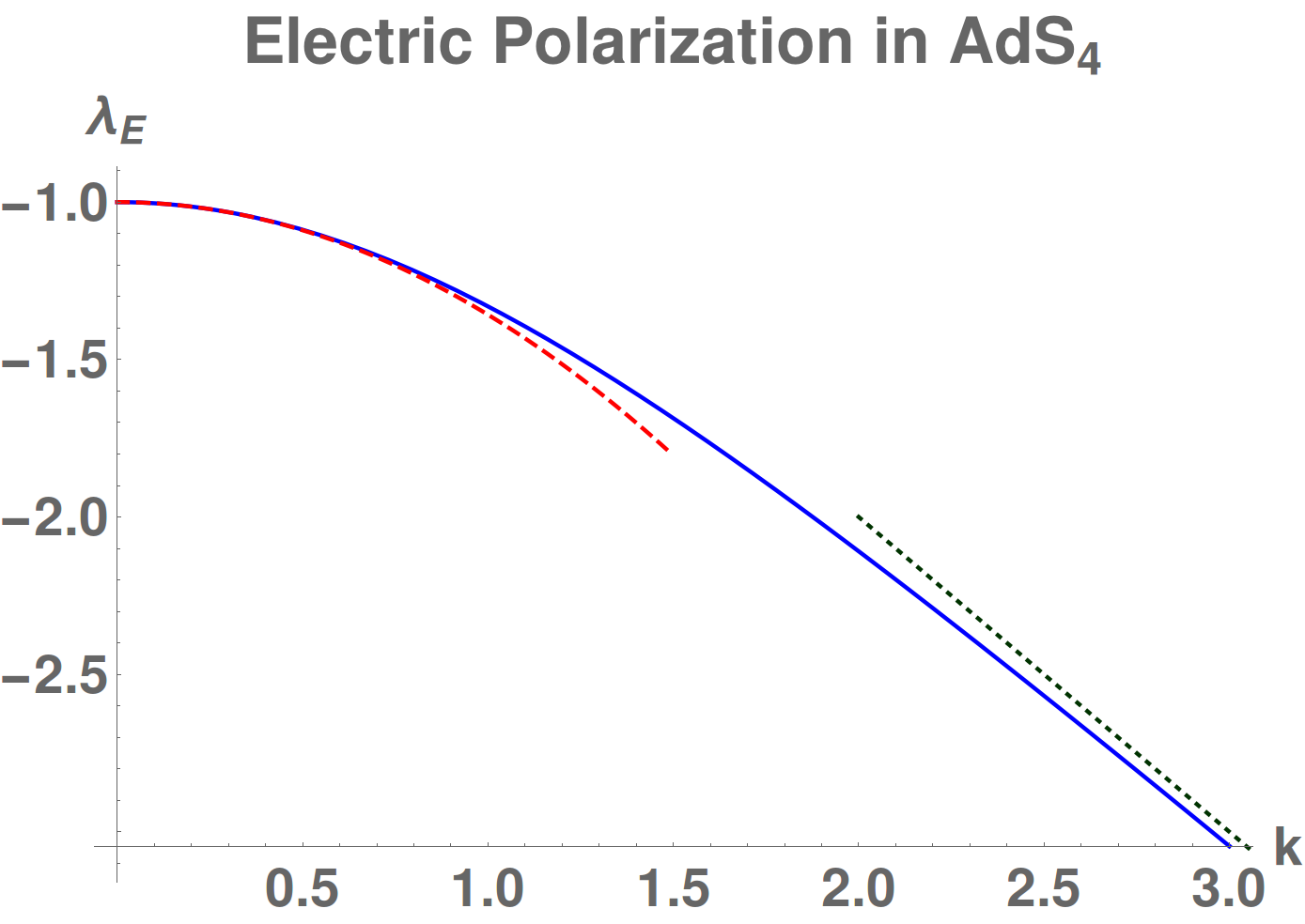}
\caption{\small Electric polarization response of black branes in AdS$_5$ and AdS$_4$ as a function of wavenumber $k$, in units $\mu=1$. Solid blue: numerical results. Dashed red: perturbative expansions in powers of $k$, eqs.~\eqref{LaE5}, \eqref{LaE4}. Dotted green: large-$k$ limit \eqref{vacel}. \label{fig:MaxwellPlots}}
\end{center}
\end{figure}

The results of the numerical and hydrodynamic evaluations of $\lambda_E(k)$ are presented in fig.~\ref{fig:MaxwellPlots}.

Observe that as $k\to 0$ the electric polarization $\lambda_E$ and the charge density $\langle J^t\rangle$ take non-zero values. This is indeed expected: this is a uniform perturbation of the black brane that adds a uniform charge distribution to it. What we then have is the Reissner-Nordstrom AdS black brane in the limit of small, linearized charge density (which does not backreact on the geometry). 

Of course this uniform charge is not a polarization effect. The way to remove it is simple. Rather than a charge density induced by an electric potential, the actual polarization effect is the charge separation  in the neutral plasma, \ie the appearance of a dipole distribution 
\beq
D_z=\partial_z J^t=\textrm{Re}(ik J^t)
\eeq
induced as a response to an external electric field 
\beq
E_z=\partial_z A_t=\textrm{Re}(ik A_t)\,.
\eeq
Then when $k\to 0$ the dipole polarization vanishes.

Notice that a similar remark could be applied to the geometric polarization: like in the Casimir effect, the measurable effect of the polarization is not so much the energy itself but the force that arises when the geometrical set up varies.

\section{Final comments}

Clearly we have only taken a first step. There is still further work ahead if one wants to test a holographic calculation of the polarization response against results from real-world systems. In particular the holographic modelling must be made more sophisticated. But we have identified the basic features of the phenomenon, and the extension to other models developed in AdS/CMT is possible.

In this article the initial unperturbed geometry for the field theory has always been Minkowski space, and correspondingly we have worked in the Poincar\'e patch of AdS in the bulk. But it is also possible and interesting to study the electric and gravitational polarization of black holes in global AdS---in dual terms, the polarization of the plasma on a spherical space. Indeed, the fully non-linear effects of electric polarization for these black holes have been studied numerically already in \cite{Costa:2015gol,Costa:2017tug}, see also \cite{Herdeiro:2015vaa,Herdeiro:2016xnp,Herdeiro:2016plq}. The analysis in global AdS is technically more complicated (spherical harmonics instead of plane waves) and presumably less relevant to systems in the lab, so we have not attempted it here.

\section*{Acknowledgments}
We are very grateful to Tom\'as Andrade, Miguel Costa, Aristos Donos, Jerome Gauntlett, Rafael Porto, Jorge Santos, and especially to Christiana Pantelidou for conversations on this subject. Some of these discussions took place during the ``Workshop on holography, black holes and numerical relativity'' at the STAG Research Centre, Southampton University, whose hospitality RE gratefully acknowledges. RE is also thankful to the Erwin Schr\"odinger Institute in Vienna for hospitality during the programme ``Quantum Physics and Gravity''.
Work supported by MEC FPA2013-46570-C2-2-P, FPA2016-76005-C2-2-P, AGAUR 2009-SGR-168, ERC Advanced Grant GravBHs-692951, MECD FPU15/01414.


\addcontentsline{toc}{section}{Appendices}
\appendix

\section{Holographic stress tensor}\label{app:cterm}

The Brown-York stress-energy tensor $\hat{T}_{\alpha\beta}$ is computed in the AdS boundary with regularized metric $\hat{g}_{\alpha\beta}$ at constant, small $v$. The renormalized metric is \eqref{renmet} and the renormalized stress-energy tensor is
\beq
T_{\alpha\beta}=\lim_{v\to 0} \lp\frac{R}{v}\rp^{n-2}\hat{T}_{\alpha\beta}\,.
\eeq

We compute it using counterterm subtraction in AdS$_4$ and AdS$_5$ ($n=3,\,4$) \cite{Balasubramanian:1999re}, in which 
\begin{equation}\label{holotab}
8\pi G\,\hat{T}_{\alpha\beta} = K_{\alpha\beta} - K\hat{g}_{\alpha\beta} -\frac{n-1}{R} \hat{g}_{\alpha\beta} + \frac{R}{n-2} \hat{G}_{\alpha\beta} -\frac{R^3}{12} \left( H^1_{\alpha\beta} - 3 H^2_{\alpha\beta}\right)\log (v e^b),
\end{equation}
where $G_{\alpha\beta}$ is the Einstein tensor of the boundary metric $\hat{g}_{\alpha\beta}$, and the last two terms, which enter only in AdS$_5$ due to the conformal anomaly, are  
\begin{equation}
H^1_{\alpha\beta} = \frac{1}{\sqrt{-\hat{g}}}\frac{\delta (\sqrt{-\hat{g}}\hat{R}^2)}{\delta \hat{g}^{\alpha\beta}} = 2\nabla_\alpha \nabla_\beta \hat{R} - 2 \hat{g}_{\alpha\beta} \nabla_\rho \nabla^\rho \hat{R} - \frac12 \hat{g}_{\alpha\beta} \hat{R}^2 +2 \hat{R} \hat{R}_{\alpha\beta},
\end{equation}
\begin{equation}\begin{split}
H^2_{\alpha\beta} = \frac{1}{\sqrt{-\hat{g}}}\frac{\delta (\sqrt{-\hat{g}}\hat{R}_{\rho\sigma}\hat{R}^{\rho\sigma})}{\delta \hat{g}^{\alpha\beta}} = & 2\nabla_\rho\nabla_\beta \hat{R}^\rho_\alpha - \nabla_\rho \nabla^\rho \hat{R}_{\alpha\beta}- \frac12 \hat{g}_{\alpha\beta}\nabla_\rho \nabla^\rho \hat{R} \\&  -  \frac12 \hat{g}_{\alpha\beta} \hat{R}_{\rho\sigma}\hat{R}^{\rho\sigma} + 2 \hat{R}_\alpha^\rho \hat{R}_{\rho \beta}.
\end{split}\end{equation}
Here all geometric quantities refer to the metric $\hat{g}_{\alpha\beta}$. The constant $b$ in \eqref{holotab} is arbitrary and reflects a renormalization scheme dependence.\footnote{Actually one can include finite contributions to the stress tensor \eqref{holotab} (and \eqref{Tmunu5}) from $H^1$ and $H^2$ with separate coefficients. For simplicity we do not do it, and our choice above is such that the stress tensor is traceless.} This ambiguity could be fixed by \eg\ imposing supersymmetry on the boundary \cite{Assel:2015nca}, but this is not particularly well motivated in our set up.

\section{Hydrodynamic expansions}\label{app:hydro}

The following are the solutions obtained in a power series expansion in $k$. They are valid for all $0<v\leq 1$. We set for simplicity $\mu=1$.

\subsection{AdS$_5$}

\subparagraph{Gravitational polarization:}

\begin{equation}\begin{split}
Z_T(v) =& 1-\frac{1}{4} \log \left(1+v^2\right) k^2+\frac{1}{128} \left(\pi ^2-4 (\log2) ^2+8 \log \left(\frac{2}{1-v^2}\right) \log \left(1+v^2\right)\right. \\& +\left. 8 \log 2 \log \left(1-v^4\right)-8 \text{Li}_2\left(\frac{1+v^2}{2}\right)-2 \text{Li}_2\left(1-v^4\right)\right) k^4+O\left(k^6\right)\,,
\end{split}\end{equation}
\begin{equation}\begin{split}
Z_V(v) =& 1-v^4-\frac{1}{4} v^2 \left(1-v^2\right) k^2 \\&+\frac{1}{32} \left(v^2(1-v^2)-2 v^4 \log v-\left(1-v^4\right) \log \left(1+v^2\right)\right) k^4+O\left(k^6\right)\,,
\end{split}\end{equation}
\begin{equation}
Z_S(v)= 1+v^4+\frac{1}{12} \left(-4 v^2\left(1+v^2\right)+\left(1+v^4\right) \log \left(1+v^2\right)\right) k^2+O\left(k^4\right)\,.
\end{equation}
These are all finite and indeed analytic functions at $v=1$.

The solutions of the constraint equations are
\begin{equation}
h_{tt}(v)=C\sqrt{1-v^4}\lp 1+v^4\rp+\frac{\bar{H}}{6} \left(1-v^2\right) \left(1-v^4\right) k^2+O\left(k^4\right)
\end{equation}
\begin{equation}
h(v)=2\bar{H}-2C\sqrt{1-v^4}+\frac{\bar{H}}{6} \left(\log \left(1+v^2\right)-2 \left(1+v^2\right)\right) k^2+O\left(k^4\right)
\end{equation}
\begin{equation}\begin{split}
h_{zz}(v)=&\bar{h}_{zz}+C\lp 1-\sqrt{1-v^4}\rp \\
&-\frac{\bar{H}}{6} \left(v^2+\log \left(1+v^2\right)-6C\arcsin (v^2) \right) k^2+O\left(k^4\right)\,.
\end{split}\end{equation}

Observe here the presence of an integration constant $C$, which corresponds to
\beq
C=\bar{h}_{tt}-\frac{\bar{H}}6k^2+O\left(k^4\right)\,.
\eeq
This constant corresponds to the gauge freedom discussed in \eqref{vchange}, \eqref{hvchange}. The gauge-invariant function $Z_S(v)$ is independent of it, but when $C\neq 0$ the metric functions $h_{tt}$, $h$, $h_{zz}$ are not analytic at the horizon position $v=1$. Therefore if we choose a gauge where the metric is analytic on the horizon, this implies that (restoring now $\mu$, and adding the next order in $k$) 
\beq\label{htt5}
\bar{h}_{tt}=\frac{\bar{H}}{6}\lp \frac{k^2}{\sqrt{\mu}}+\frac{k^4}{24\mu}\lp \pi-12+6\log 2\rp\rp + O\left(k^6\right)\,.
\eeq

\subparagraph{Electric polarization:}
\begin{equation}\begin{split}
Z_E(v) = 1-v^2+\frac{1}{4} \left(2v^2 \log \left(2v\right)-\left(1+v^2\right) \log \left(1+v^2\right)\right) k^2+O\left(k^6\right)\,.
\end{split}\end{equation}

\subsection{AdS$_4$}

\subparagraph{Gravitational polarization:}

\begin{equation}\begin{split}
Z_V(v) =& 1-v^3-\frac{1}{2} (1-v) v^2 k^2 \\& -\frac{1}{108} \left(9 v (1-v)(2+v)+2 \sqrt{3} \left(1-v^3\right)\lp \pi  -6\arctan\left(\frac{1+2 v}{\sqrt{3}}\right)\rp\right) k^4+O\left(k^6\right)
\end{split}\end{equation}
\begin{equation}\begin{split}
Z_S(v)= 1+\frac{v^3}{2} -\frac{v^2 k^2}{2}+ \frac{1}{216} \Bigg(& 36 v\left(1+v^2\right)+\sqrt{3}   \left(2+v^3\right)\lp\pi-6\arctan\left(\frac{1+2 v}{\sqrt{3}}\right)\rp \\&  -9 \left(2+v^3\right) \log \left(1+v+v^2\right)\Bigg) k^4+O(k^6)\,,
\end{split}\end{equation}
with metric functions
\begin{equation}\begin{split}
h_{tt}(v)= \frac{\bar{H}}{24}k^2 \Bigg(&-4v^2(1- v^3)+\frac{\left(2+v^3\right) \sqrt{\pi(1-  v^3)}\, \Gamma \left(\frac{5}{3}\right)}{\Gamma \left(\frac{7}{6}\right)}  \\&+ v^2 \left(-2+v^3+v^6\right) \, _2F_1\left(1,\frac{7}{6};\frac{5}{3};v^3\right)\Bigg) +O\left(k^4\right)\,,
\end{split}\end{equation}
\begin{equation}
h_{xx}(v)= \bar{H}\lp 1-\frac{k^2}{12}\left(\frac{\sqrt{\pi(1 - v^3)}\, \Gamma \left(\frac{5}{3}\right)}{\Gamma \left(\frac{7}{6}\right)}+ v^2 \left(4-\left(1-v^3\right) \, _2F_1\left(1,\frac{7}{6};\frac{5}{3};v^3\right)\right)\right) \rp+O\left(k^4\right)\,,
\end{equation}
\begin{equation}
h_{zz}(v)=\bar{h}_{zz} +\bar{H}\left(\frac{\sqrt{\pi } \left(1-\sqrt{1-v^3}\right)\, \Gamma \left(\frac{5}{3}\right)}{12 \Gamma \left(\frac{7}{6}\right)}-\frac{1}{40} v^2 \left(10+v^3 \, _2F_1\left(1,\frac{7}{6};\frac{8}{3};v^3\right)\right)\right) k^2+O\left(k^4\right)\,.
\end{equation}

Since the expressions are cumbersome, here we have already chosen the analytic gauge, which determines (now with $\mu$ restored)
\begin{equation}\label{htt4}
\bar{h}_{tt}(k) = \bar{H}\lp\frac{k^2}{\mu^{2/3}} \frac{\sqrt{\pi }\, \Gamma \left(\frac{5}{3}\right)}{12 \Gamma \left(\frac{7}{6}\right)}-\frac{k^4 }{9\mu^{4/3}}\left(1-\frac{\sqrt{3} \pi ^{3/2}}{9\Gamma
   \left(\frac{2}{3}\right) \Gamma \left(\frac{5}{6}\right)}\right)\rp + O(k^6).
\end{equation}

\subparagraph{Electric polarization:}
\begin{equation}\begin{split}
Z_E(v) = 1 - v+\frac{k^2}{2} \Biggl(&\frac{2 (2 v+1) }{\sqrt{3}}\arctan\left(\frac{2
   v+1}{\sqrt{3}}\right)-\frac{\pi  (5 v+1)}{3 \sqrt{3}}\\ &+v \log 3-\log \left(v^2+v+1\right)\Biggr)+ O(k^4)\,.
\end{split}\end{equation}

\section{Analytic gauge}\label{app:angauge}

In the main text we have discussed that certain choices of the radial coordinate $v$ lead to metric functions $h_{tt}(v)$, $h(v)$, $h_{zz}(v)$ that behave like $\sim \sqrt{1-\mu v^n}$ near the horizon at $v=\mu^{-1/n}$. This non-analyticity is inconvenient for showing that the horizon is regular. For instance, if one changes $(t,v)\to (x^+,v)$ where the latter are ingoing Eddington-Finkelstein coordinates, then if the $v$-gauge is not analytic the metric in these coordinates is singular at the horizon. Proving horizon regularity requires to first perform a change of the type \eqref{vchange} to an analytic radial gauge. Nevertheless, invariants such as the surface gravity can be computed in any radial gauge.

\begin{figure}[t]
\begin{center}
\includegraphics[width=200pt]{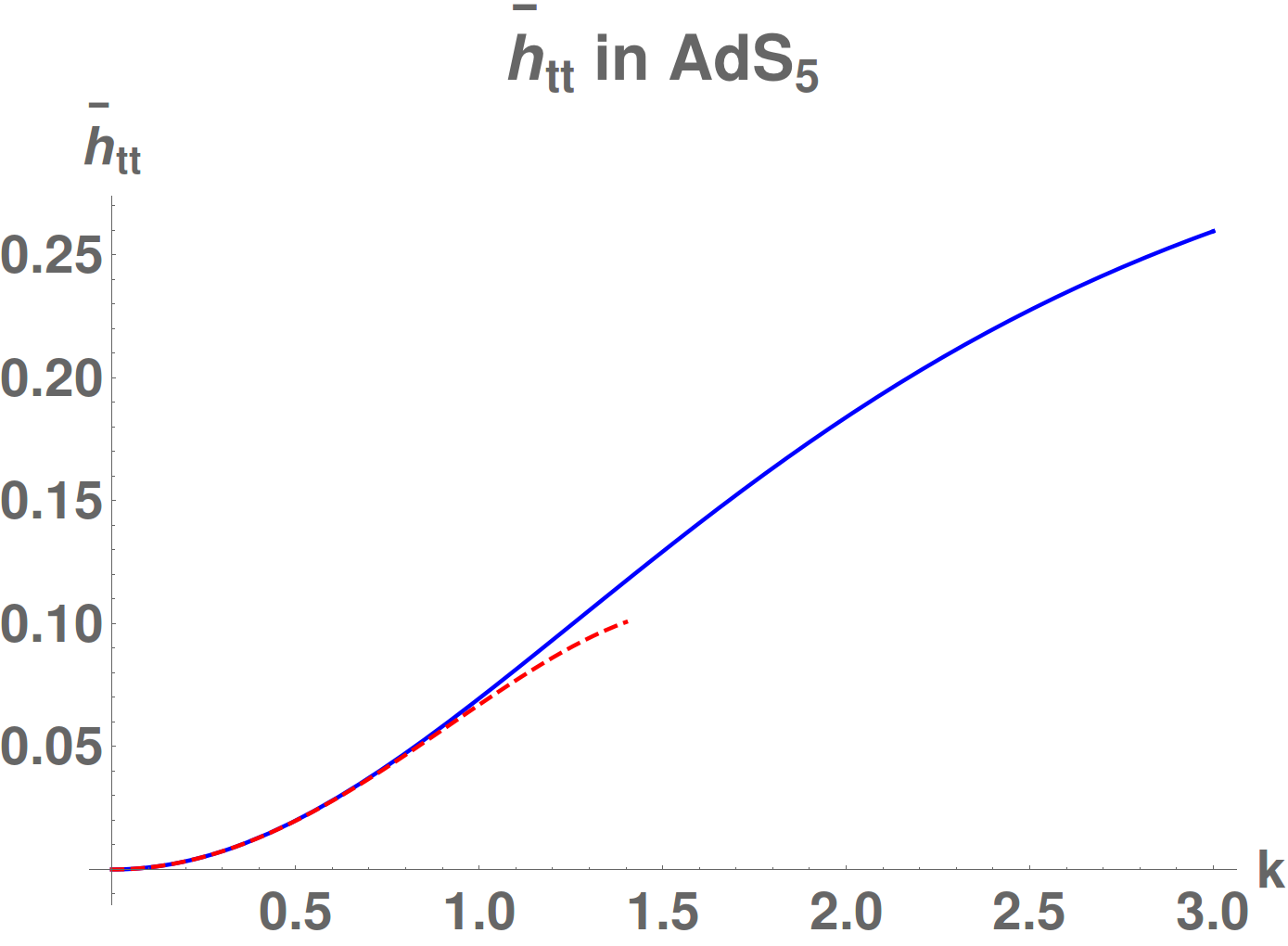}\qquad
\includegraphics[width=200pt]{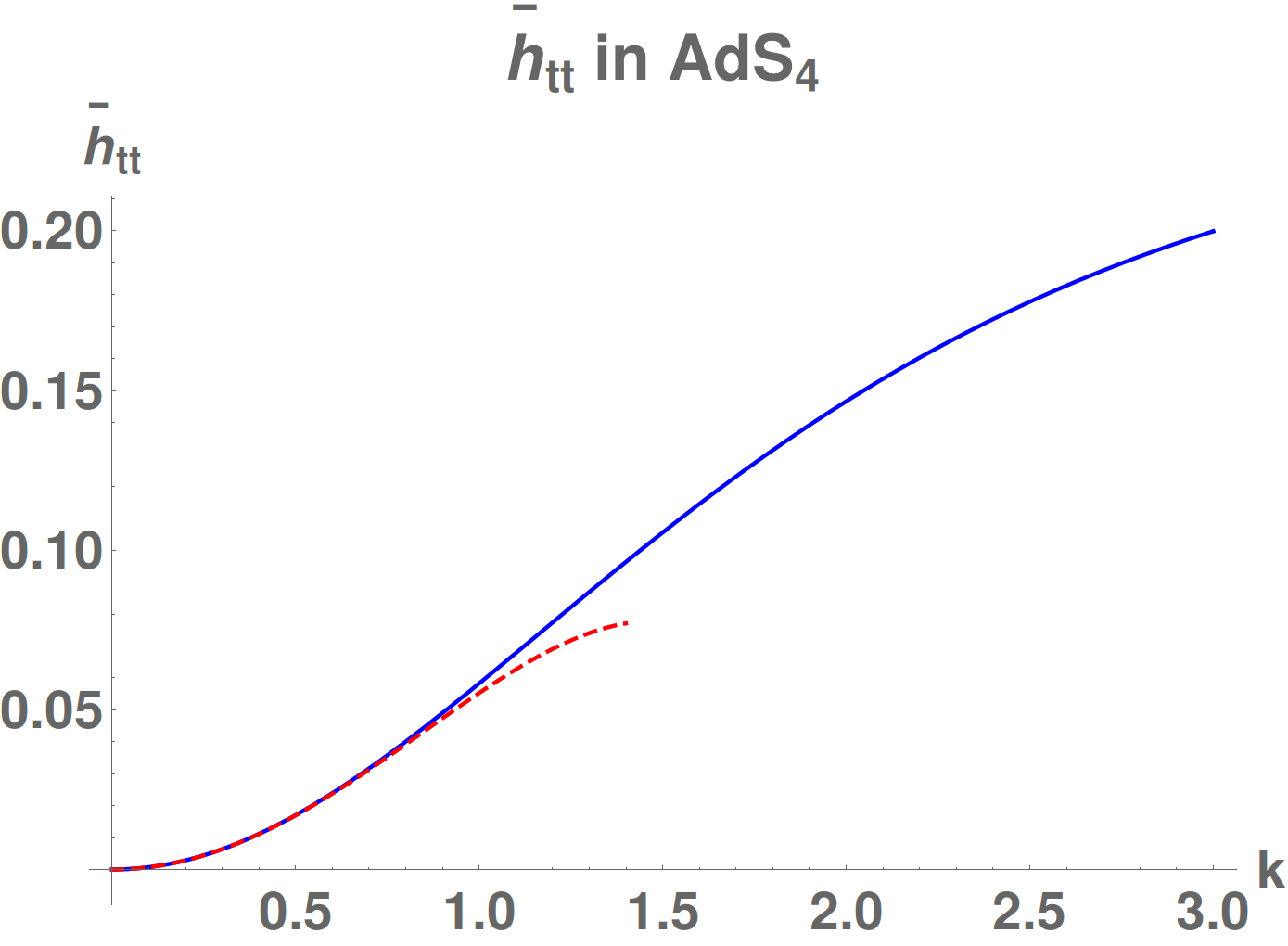}
\caption{\small Values of $\bar{h}_{tt}(k)$ in the analytic gauge, for AdS$_5$ and AdS$_4$. Solid blue: numerical results. Dashed red: perturbative expansions in powers of $k$, eqs.~\eqref{htt5}, \eqref{htt4}. \label{fig:BetaPlots}}
\end{center}
\end{figure}

The transformations \eqref{vchange} alter $\bar{h}_{tt}$. Fig.~\ref{fig:BetaPlots} gives the values of $\bar{h}_{tt}(k)$ that result when taking the analytic gauge. We compare them with the hydrodynamic calculations of appendix~\ref{app:hydro}.

\end{document}